\documentclass[fleqn,usenatbib,useAMS]{mnras}

\usepackage[dvipdfmx]{graphicx}	% Including figure files
\usepackage{amsmath}	% Advanced maths commands
\usepackage{amssymb}	% Extra maths symbols
\usepackage{multicol}        % Multi-column entries in tables
\usepackage{bm}		% Bold maths symbols, including upright Greek
\usepackage{pdflscape}	% Landscape pages
\usepackage{ulem}

\newcommand{\msun}{M_\odot}
\newcommand{\rsun}{R_\odot}

\newcommand{\const}{\mathrm{const.}}

\newcommand{\revise}[1]{{\textcolor{black}{#1}}}

\usepackage[T1]{fontenc}
\usepackage{ae,aecompl}

\usepackage{newtxtext,newtxmath}
\title[Pop. III BBH mergers beyond the mass gap]{On the population III binary black 
hole mergers beyond the pair-instability mass gap}

\author[Kotaro Hijikawa]{
Kotaro Hijikawa,$^{1}$\thanks{E-mail: hijikawa@astron.s.u-tokyo.ac.jp}
Ataru Tanikawa,$^{2}$ Tomoya Kinugawa,$^{3}$ Takashi Yoshida$^{4,1}$ and
Hideyuki Umeda$^{1}$
\\
$^{1}$Department of Astronomy, Graduate School of Science, 
The University of Tokyo, 7-3-1 Hongo, Bunkyo-ku, 113-0033 Tokyo, Japan\\
$^{2}$Department of Earth Science and Astronomy, College of
Arts and Sciences, The University of Tokyo, Meguro-ku, Tokyo, Japan\\
$^{3}$Institute for Cosmic Ray Research, The University of Tokyo,
Kashiwa, Chiba, Japan\\
$^{4}$Yukawa Institute for Theoretical Physics, Kyoto University, Kitashirakawa Oiwakecho, Sakyo-ku, Kyoto 606-8502, Japan}

\date{Accepted XXX. Received YYY; in original form ZZZ}

\pubyear{2021}

\begin{document}
\label{firstpage}
\pagerange{\pageref{firstpage}--\pageref{lastpage}}
\maketitle

\begin{abstract}
We perform a binary population synthesis calculation incorporating
very massive population (Pop.) III stars up to 1500 $\msun$, and
investigate the nature of binary black hole (BBH) mergers. 
Above the pair-instability mass gap, we find that the typical primary 
black hole (BH) mass is 135--340 $\msun$. 
The maximum primary BH mass is as massive as 686 $\msun$. 
The BBHs with both of their components above the mass gap have low 
effective inspiral spin $\sim$ 0.  
So far, no conclusive BBH merger beyond the mass gap has been detected, 
and the upper limit on the merger rate density is obtained. 
If the initial mass function (IMF) of Pop. III stars is simply expressed 
as $\xi_m(m) \propto m^{-\alpha}$ (single power law), we find that 
$\alpha \gtrsim 2.8$ is needed in order for the merger rate density not 
to exceed the upper limit.
In the future, the gravitational wave detectors such as Einstein 
telescope and Pre-DECIGO will observe BBH mergers at high redshift. 
We suggest that we may be able to impose a stringent limit on the Pop. 
III IMF by comparing the merger rate density obtained from future 
observations with that derived theoretically.
\end{abstract}


%%%%%%%%%%%%%%%%%%%%%%%%%%%%%%%%%%%%%%%%%%%%%%%%%%

%%%%%%%%%%%%%%%%% BODY OF PAPER %%%%%%%%%%%%%%%%%%

% The MNRAS class isn't designed to include a table of contents, but for this document one is useful.
% I therefore have to do some kludging to make it work without masses of blank space.
%\begingroup
%\let\clearpage\relax
%\tableofcontents
%\endgroup
%\newpage

\section{Introduction}
The gravitational wave (GW) observations have been conducted by the GW
observatories, the advanced laser interferometer gravitational wave
observatory (aLIGO) and advanced VIRGO, and the new GW catalog,
GWTC-2, was recently released by the LIGO Scientific Collaboration and 
the Virgo Collaboration \citep{abbott20gwtc2}.  This catalog includes 50 
GW sources and 47 of which are binary black holes (BBHs). 
The typical mass of the detected BBHs is $\sim$ 30 $\msun$ , and this 
is heavier than that of X-ray binary BHs.

Origins of BBH mergers have been proposed by many authors: field
population (Pop.) I or II binary stars \citep[e.g.,][]{belczynski20,
kruckow18}, dynamical interactions in star clusters
\revise{\citep[e.g.,][]{rodriguez16, dicarlo20, kumamoto20}} , AGN disks
\citep[e.g.,][]{tagawa20}, and triple or multiple star systems
\citep[e.g.,][]{antonini17}.  Pop. III binary stars can also become
BBH mergers.  The typical mass of Pop. III BBHs is $\sim$ 30 $\msun$ +
30 $\msun$ \citep[e.g.][]{kinugawa14,kinugawa20}, and this is
consistent with the observation \citep{kinugawa21}.  Therefore,
Pop. III binary stars are one of the promising origins of BBH mergers.

Previously, \citet{belczynski04}, \citet{kinugawa14,kinugawa20} and 
\citet{tanikawa21} performed binary population synthesis calculations 
for Pop. III stars and predicted the property of BBH mergers.
The maximum zero-age main-sequence (ZAMS) masses in their calculations 
are 500 $\msun$, 150 $\msun$ and 300 $\msun$, respectively.
According to Pop. III star formation simulations 
\citep[e.g.,][]{susa14, hirano15}, there is a chance that Pop. III stars
as massive as $\sim$ 1000 $\msun$ could be formed 
\citep[e.g.,][]{hirano15}, although the typical mass is 10--100 $\msun$.
Future GW observatories, such as Cosmic Explorer \citep{reitze19}, 
Einstein telescope \citep{punturo10,sathyaprakash2012} and Pre-DECIGO 
\citep{kawamura2006,nakamura2016} can detect very massive compact binary 
mergers (total mass $\sim$ 100--1000 $\msun$) up to the redshift $z\sim$ 
10--1000.
If there are BBHs originating from very massive Pop. III stars, 
these observatories may be able to detect them.
Therefore, in this Letter, we focus on the very massive Pop. III stars 
up to 1500 $\msun$ and investigate the contribution of very massive Pop.
III stars to the BBH merger by means of binary population synthesis in 
preparation for future observations.

\section{Method}\label{sec:method}
\subsection{Binary Population Synthesis}
In order to perform a population synthesis calculation, we use the
binary population synthesis code \citep{tanikawa21}, which is an
upgraded version of {\tt BSE} \citep{hurley00,hurley02}
\footnote[1]{\url{http://astronomy.swin.edu.au/\~jhurley/}}.
In our code, the fitting formulae for extremely metal poor stars were
implemented \revise{\citep{tanikawa20a,tanikawa21}}. 
We use the same mass transfer rate for the stable Roche-lobe overflow 
and tidal coefficient factor $E$ as in \citet{kinugawa20}.

We use the $\alpha\lambda$ formalism for common envelope evolution
\citep{webbink84} and set $\alpha_\mathrm{CE}=1$ and
$\lambda_\mathrm{CE}=1$.
We adopt the `rapid' model in \citet{fryer12} with the modification for 
the pulsational pair-instability and pair-instability supernova 
\citep[see equations 5--7 in][]{tanikawa21}.
We assume that the mass of a BH formed through pulsational 
pair-instability is 45 $\msun$\revise{, and the minimum helium core mass 
of direct collapse is 135 $\msun$ in the same way as \citet{belczynski16b}.
}

\subsection{Initial Conditions}
In this study, we follow the evolution of $10^6$ binaries consisting
of ZAMS stars with the metallicity $Z=10^{-8}Z_\odot$.
\revise{Although we adopt the metallicity $Z=10^{-8}Z_\odot$ for
  Pop. III stars ($Z=0$), this difference has little effect on our
  results for the following reason.}  \revise{As the abundance of
  heavy elements in a $Z=10^{-8}Z_\odot$ star is extremely low, the
  CNO cycle does not operate initially. The abundance of carbon,
  nitrogen and oxygen are increased by the triple alpha reaction, and
  then the CNO cycle commences. These evolutionary properties are the
  same as those of $Z=0$ stars. Therefore, the difference in
  metallicity, $Z=0$ and $Z=10^{-8}Z_\odot$, has little effect on the
  stellar evolution.  }

We use the following initial distributions for the primary mass, mass 
ratio, and orbital separation.
The initial primary mass $m_\mathrm{ZAMS,p}$ distribution is logarithmically flat
($\xi_m(m_\mathrm{ZAMS,p}) \propto m_\mathrm{ZAMS,p}^{-1}$), and the mass range is 10--1500
$\msun$.
The initial mass ratio $q$ distribution is flat 
($\xi_q(q)\propto\const$), and the range is $q_\mathrm{min}$--1, where
$q_\mathrm{min}=10\msun/m_\mathrm{ZAMS,p}$.
The initial orbital separation $a$ distribution is logarithmically flat,
and the range is $a_\mathrm{min}$--$10^5\rsun$, where $a_\mathrm{min}$ 
is determined so that the ZAMS radius does not exceed the Roche lobe 
radius.
In this study, we assume that the initial eccentricity is zero.

We adopt the star formation history in \citet{desouza11} for Pop. III
stars, but reduce it by a factor of three
\citep{inayoshi16,kinugawa20}.
We assume that the binary fraction $f_\mathrm{b}$ is 0.5.

\section{Results}
\subsection{Mass Distribution}
\begin{figure}
   \begin{center}
      \includegraphics[width=\columnwidth]{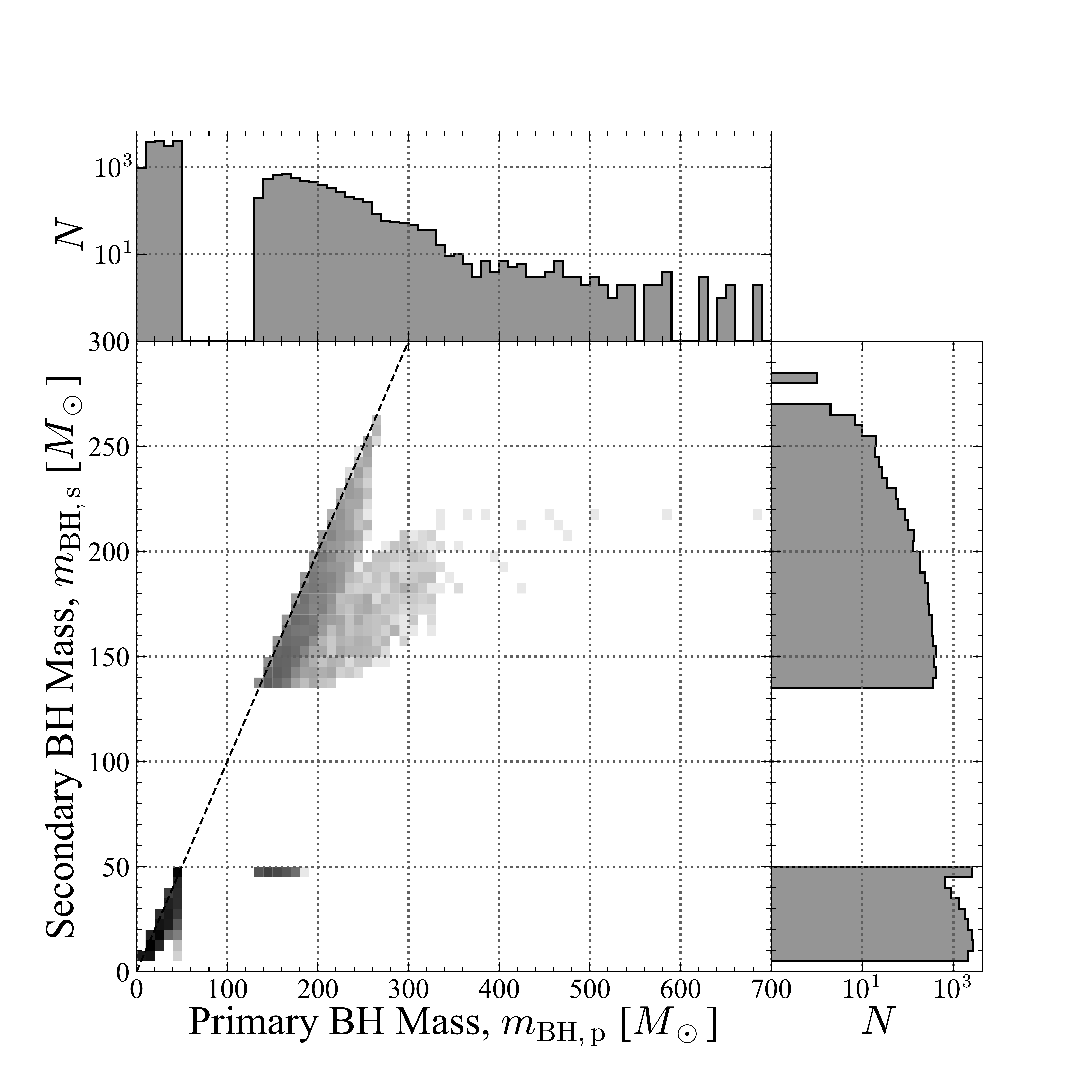}
   \end{center}
   \caption{Mass distribution of the BBHs which merge within a 
   Hubble time. The dashed black line indicates the line where primary 
   mass $m_\mathrm{BH,p}$ is equal to the secondary mass 
   $m_\mathrm{BH,s}$.
	Each 1D histogram and the shades of color for 2D histogram are on a 
   log-scale.
	\label{fig:mass_dist}}
\end{figure}

The mass distribution of BBHs which merge within a Hubble time is shown
in Figure \ref{fig:mass_dist}.
The distribution is divided into three sub-populations due to the 
pair-instability mass gap.
Hereafter, we call the combination of low mass BH ($\lesssim$ 50 
$\msun$) and low mass BH `low mass + low mass', that of low mass BH and 
high mass BH ($\gtrsim$ 130 $\msun$) `low mass + high mass', and that of 
high mass BH and high mass BH `high mass + high mass'.

\begin{figure}
   \begin{center}
      \includegraphics[width=\columnwidth]{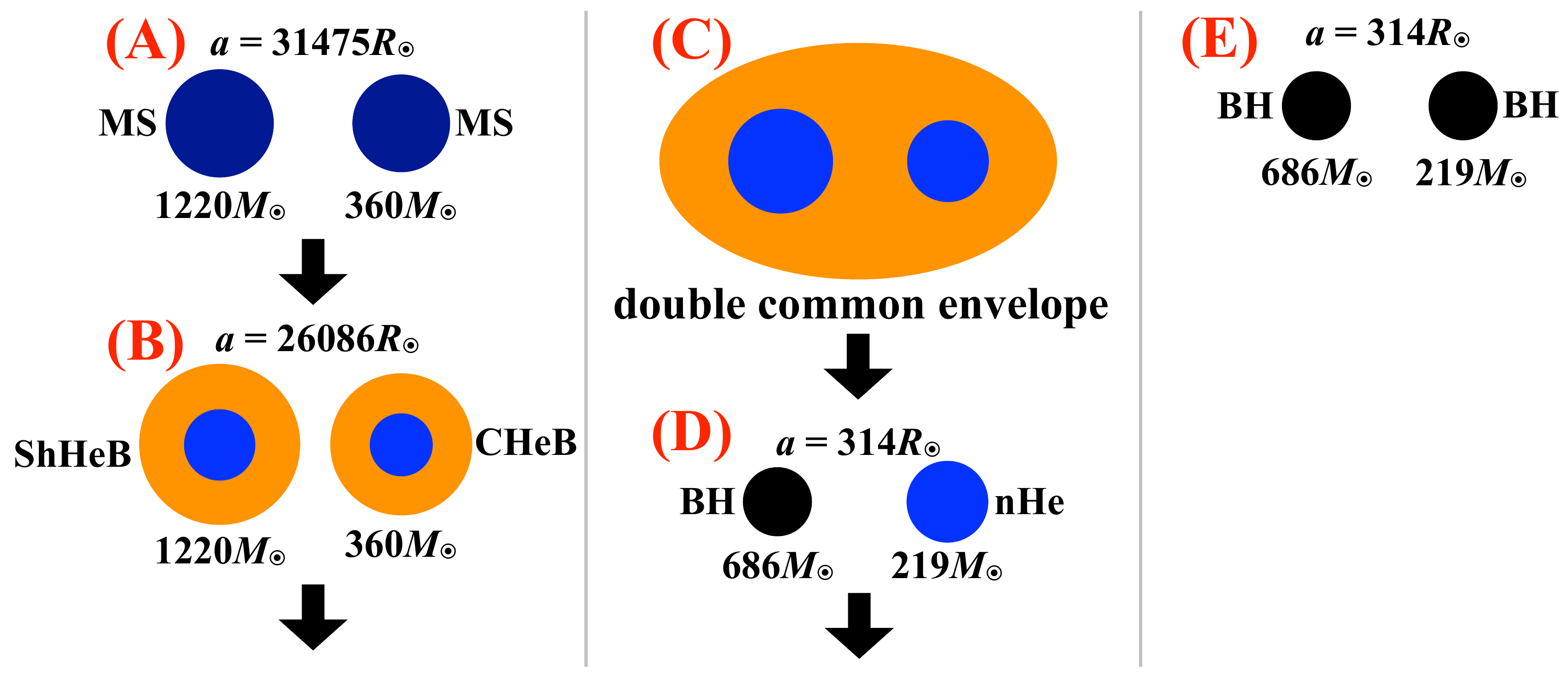}
   \end{center}
   \caption{Evolutionary channel leading to the BBH with the most 
   massive primary BH. `CHeB', `ShHeB' and `nHe' stand for the core 
   helium burning phase, shell helium burning phase and naked helium 
   star, respectively.
   \label{fig:channel_hh}}
\end{figure}
The peak of high mass primary BH is around 140--180 $\msun$.  The BBH
with the most massive primary BH in our calculation is the pair of 686
$\msun$ BH and 219 $\msun$ BH, and their ZAMS masses are 1220 $\msun$
and 360 $\msun$, respectively.  We show in Figure \ref{fig:channel_hh}
the evolutionary channel leading to this BBH.  This BBH is formed
through double common envelope scenario in which both of the stars
are giant phases with a clear core--envelope structure.  The primary ZAMS stars having a 
ZAMS mass higher than $\sim$ 1220 $\msun$ cannot contribute to the BBH 
merger.
%If the primary ZAMS mass is higher than this, it cannot contribute to the BBH merger. 
The reason for this is as follows.  
A massive Pop. III star with a ZAMS mass higher than $\sim$ 600 $\msun$ 
extremely expands,
%If the ZAMS mass of a Pop. III star is higher than $\sim600\msun$, it extremely expands, 
%becomes a red supergiant,
can reach the Hayashi track during the main sequence (MS) phase,
and has a convective envelope when it is still in its MS phase.  Therefore, if such a massive Pop. III MS star fills its
Roche lobe, the mass transfer may become unstable, and the binary
system enters a common envelope phase.  After that, the binary system
always coalesces because the MS star does not have a clear entropy
jump between its core and envelope, and it will no longer evolve to a
BBH.  In order to avoid this common envelope episode, the orbital
separation needs to be large enough not to fill the Roche lobe during
the MS phase. 

Hereafter, in order to understand why there is an upper limit on the 
primary ZAMS mass of merging BBHs, we consider the dependence of the
merging timescale of the binary system on the primary ZAMS mass.
The merging timescale through GW emission is expressed as
$t_\mathrm{GW}\propto a^4m_\mathrm{BH,p}^{-2}$, where $a$ is the
orbital separation.  Based on the above discussion, $a\propto
r_\mathrm{giant,p}$, where $r_\mathrm{giant,p}$ is the primary radius in
the giant phase.  $r_\mathrm{giant,p}\propto m_\mathrm{ZAMS,p}^{0.6}$ for
Pop. III stars with $m_\mathrm{ZAMS,p}\gtrsim600$ $\msun$, and
$m_\mathrm{BH,p}\propto m_\mathrm{ZAMS,p}$.%, where $m_\mathrm{ZAMS,p}$ is the primary ZAMS mass.  
Therefore, $t_\mathrm{GW}\propto m_\mathrm{ZAMS,p}^{0.4}$. 
As the primary ZAMS mass increases, the mering timescale increases, and exceeds 
a Hubble time at a critical ZAMS mass.
%%  Thus, the orbital separation ($a$) should be larger than the
%%  primary radius in the giant phase ($r_\mathrm{giant}$); $a \propto
%%  r_\mathrm{giant}$.  Pop. III stars with $m_\mathrm{ZAMS} \gtrsim
%%  600$ $\msun$ have $r_\mathrm{giant} \propto
%%  m_\mathrm{ZAMS}^{0.6}$, where $m_\mathrm{ZAMS}$ is the ZAMS
%%  mass. Then, the orbital separation should be $a \propto
%%  m_\mathrm{ZAMS}^{0.6}$. The merging timescale through GW emission
%%  can be expressed as $t_\mathrm{GW} \propto a^4
%%  m_\mathrm{BH,p}^{-2} \propto m_\mathrm{ZAMS}^{0.4}$, where
%%  $m_\mathrm{BH,p} \propto m_\mathrm{ZAMS}$. Thus, the merging
%%  timescale increases with the ZAMS mass increasing, and exceeds a
%%  Hubble time at a critical ZAMS mass.
%However, if the initial orbital separation is too large, it is
%difficult to merge within a Hubble time.  This is because even if the
%binary system enters a common envelope phase during a core helium
%burning phase, it cannot get close enough.  \revise{({\bf AT:} There
%seems a gap in this logic .)}
%That is why a Pop. III star with the ZAMS mass $\gtrsim$ 1220 $\msun$
%cannot contribute to the BBH mergers.  
That is why there is an upper limit on the primary ZAMS mass of merging 
BBHs.
If we use a smaller (larger)
$\alpha_\mathrm{CE}\lambda_\mathrm{CE}$, the maximum mass of merging
BBHs is expected to get larger (smaller), because how much the orbital
separation shrinks due to the common envelope friction depends on
$\alpha_\mathrm{CE}\lambda_\mathrm{CE}$.

As can be seen from Figure \ref{fig:mass_dist}, the number of the 
primary BH mass decreases sharply at $\sim$ 340 $\msun$, which is equal 
to the helium core mass of a 600 $\msun$ Pop. III ZAMS star. 
This reason is as follows.
In our calculation, all `high mass + high mass' BBHs are formed through
double common envelope phase, and some of them undergo a stable mass
transfer before the double common envelope phase.
As mentioned earlier, a Pop. III star with the ZAMS mass $\gtrsim$ 600 
$\msun$ becomes a red supergiant when it is still in its MS phase.
Therefore, if the primary ZAMS mass is $\gtrsim$ 600 $\msun$, the mass 
transfer may become unstable, and thus the binary systems coalesce, 
because the primary star does not have a clear entropy jump between its 
core and envelope.
The binary systems can not form BBHs.

\begin{figure}
   \begin{center}
      \includegraphics[width=\columnwidth]{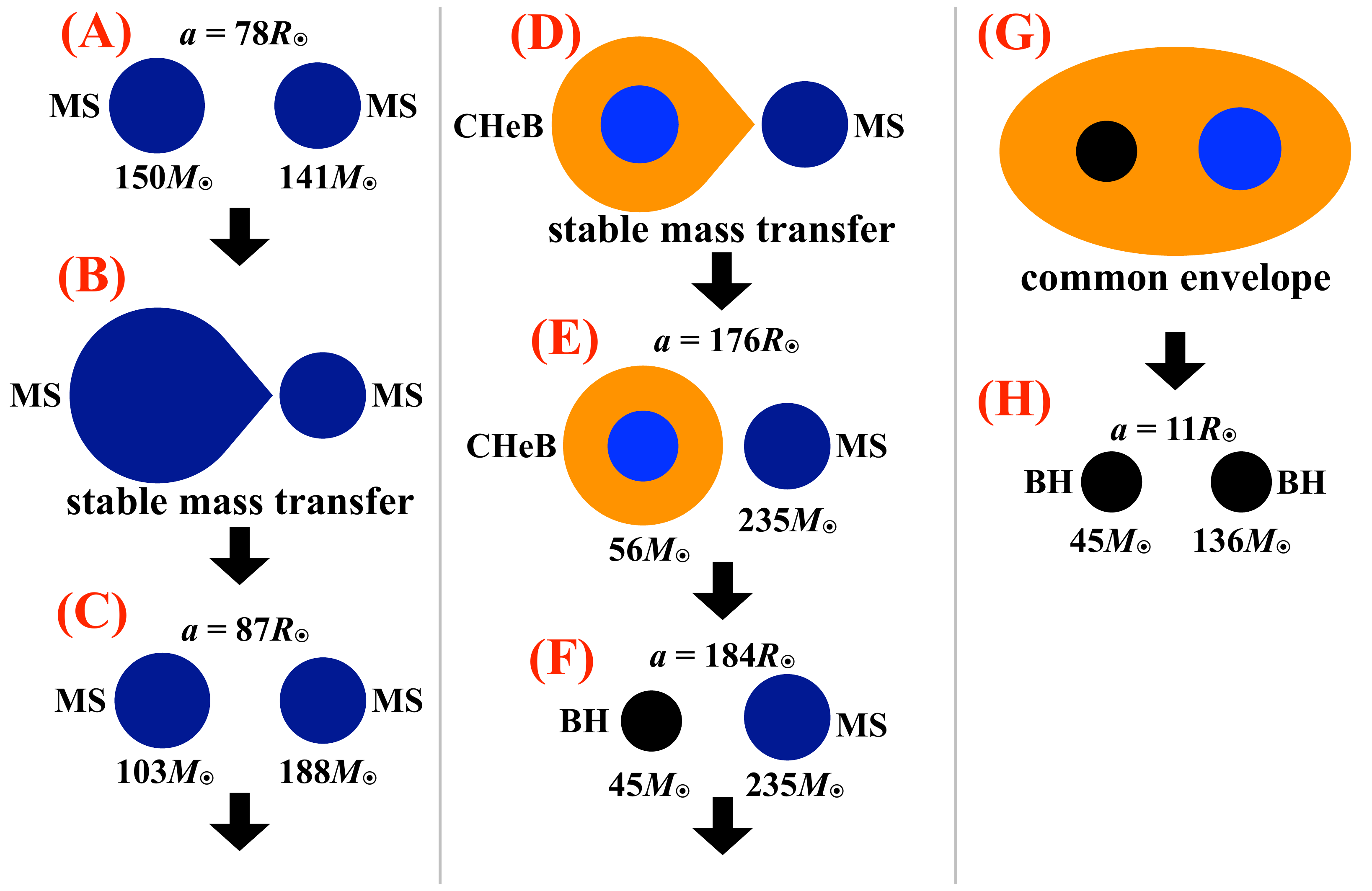}
   \end{center}
   \caption{An example evolutionary channel leading to `low mass + high 
   mass'.
   \label{fig:channel_lh}}
\end{figure}
In our calculation, all `low mass + high mass' BBHs have $\sim$ 45
$\msun$ secondary BHs, and those BHs are formed through pulsational
pair-instability.
An example evolutionary path to `low mass + high mass' is shown in 
Figure \ref{fig:channel_lh}.
The primary ZAMS mass is $\sim150\msun$, and it will generally cause a
pair-instability supernova without mass loss through mass transfer.
However, the mass transfer rate is so high that the primary star 
significantly loses its mass to the secondary star, and can avoid a 
pair-instability supernova.

\subsection{Spin Distribution}
\begin{figure}
   \begin{center}
      \includegraphics[width=\columnwidth]{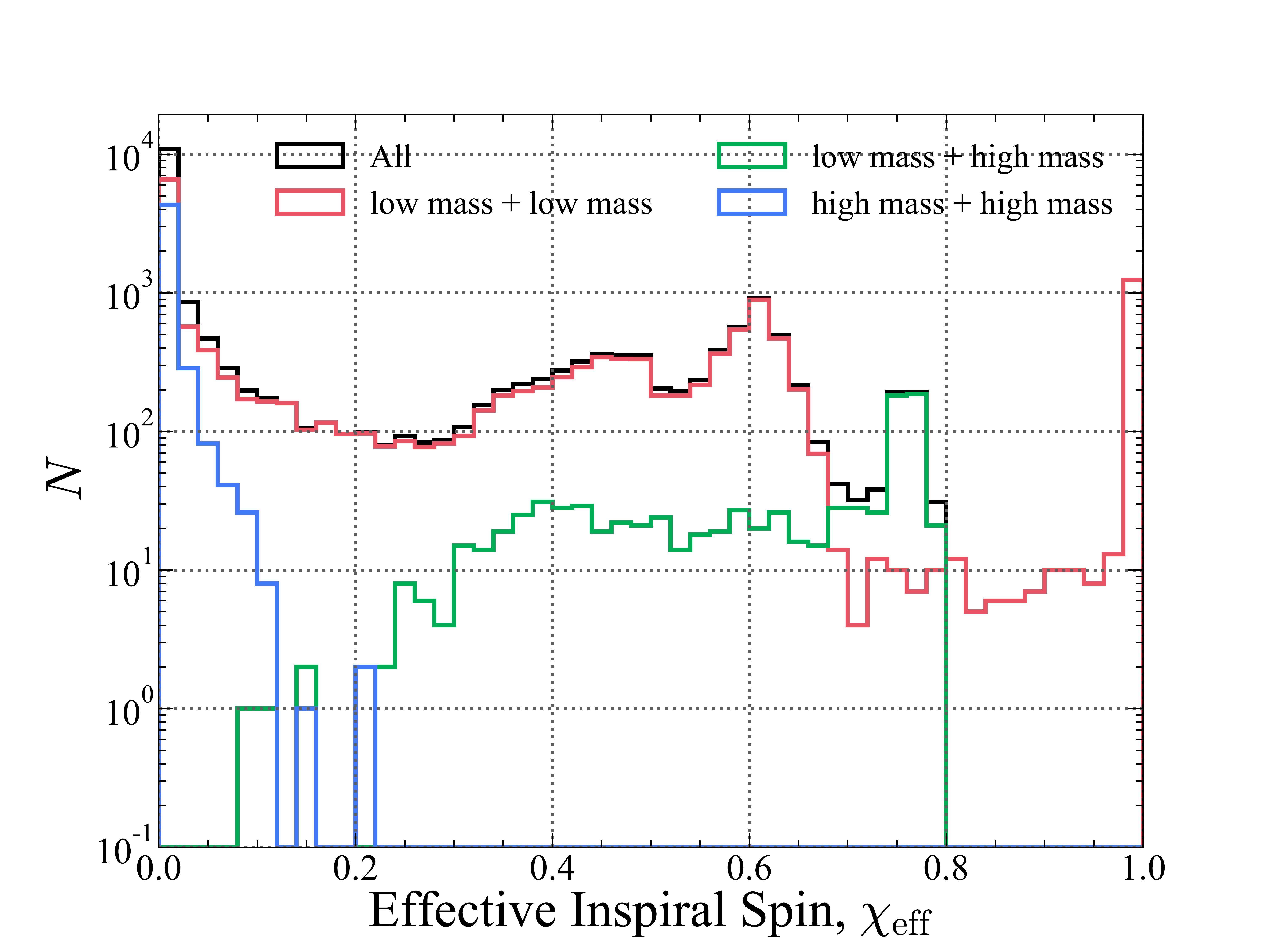}
   \end{center}
   \caption{Effective inspiral spin parameter distribution of the 
   BBHs which merge within a Hubble time. The black solid line indicates 
   the effective inspiral spin distribution of all the BBHs. The red, 
   green, and blue lines correspond to the `low mass + low mass', 
   `low mass + high mass', and `high mass + high mass', respectively. 
   \label{fig:chieff}}
\end{figure}
In Figure \ref{fig:chieff}, we show the effective inspiral spin
parameter $\chi_\mathrm{eff}$ distribution.
Since we do not assume the kick, no negative $\chi_\mathrm{eff}$ is 
obtained.

In our calculation, all `high mass + high mass' BBHs are formed
through double common envelope phase.
If a binary system enters a double common envelope phase, both stars
lose their hydrogen envelopes and spin angular momenta.
Therefore, both dimensionless spins of `high mass + high mass' are very 
low, and 99.8 \% of `high mass + high mass' BBHs have 
$\chi_\mathrm{eff}$ less than 0.1.

On the other hand, $\chi_\mathrm{eff}$ of `low mass + high mass' peaks
at around 0.75--0.80.
Since the secondary BH progenitor of `low mass + high mass' 
significantly loses its mass to the primary BH progenitor due to stable 
mass transfer (the phase B and D in Figure \ref{fig:channel_lh}), the 
secondary spin $\chi_\mathrm{BH,s}$ of all `low mass + high mass' is 
about 0.
Thanks to the short orbital separation after the common envelope phase, 
the primary BH progenitor can be highly spun up by tidal interaction.
Therefore, $\chi_\mathrm{BH,p}$ peaks at around 1, even though the 
primary BH progenitor loses its hydrogen envelope and angular momenta 
during the common envelope phase.
When the primary spin is 1, secondary spin is 0, secondary mass is 45 
$\msun$, and both spin vectors are aligned with the orbital angular 
momentum vector,
\begin{align}
   \chi_\mathrm{eff}=\frac{m_\mathrm{BH,p}\chi_\mathrm{BH,p}+m_\mathrm{BH,s}\chi_\mathrm{BH,s}}
   {m_\mathrm{BH,p}+m_\mathrm{BH,s}}=
   \frac{m_\mathrm{BH,p}}{m_\mathrm{BH,p}+45}.
\end{align}
Since the primary BH mass of `low mass + high mass' is 135--180 $\msun$,
$\chi_\mathrm{eff}=$ 0.75--0.80.
Therefore, $\chi_\mathrm{eff}$ of `low mass + high mass' peaks at around
0.75--0.80.

\subsection{Merger Rate Density}
\begin{figure}
   \begin{center}
      \includegraphics[width=\columnwidth]{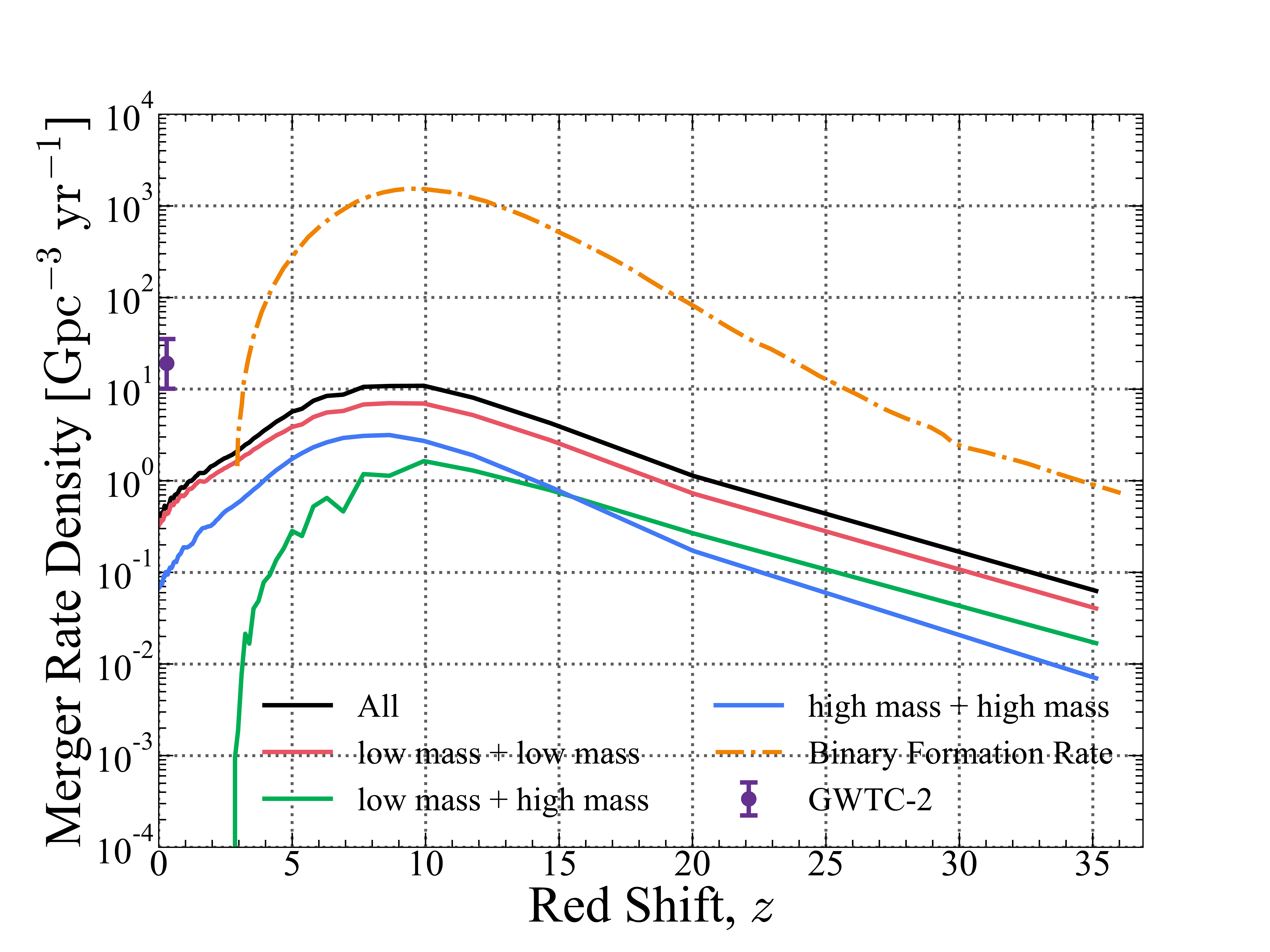}
   \end{center}
   \caption{Redshift evolution of merger rate density \revise{when the IMF
   is $\xi_m(m_\mathrm{ZAMS,p})\propto m_\mathrm{ZAMS,p}^{-1}$}. 
   The color codes are the same as Figure \ref{fig:chieff}. 
   The purple circle with the error bar indicates the merger rate 
   density estimated from GW observation \citep{abbott20gwtc2Property} 
   in the current universe.
   The orange dash-dot curve indicates the redshift evolution of the 
   binary formation rate density in number, based on \citet{desouza11}.
	\label{fig:mergerrate}}
\end{figure}

The merger rate density evolution is shown in Figure 
\ref{fig:mergerrate}.
In the current universe ($z=0$), the merger rate density of all Pop. 
III BBHs and `high mass + high mass' BBHs are 
$\mathcal{R}^\mathrm{all}(z=0)=$ 0.383 $\mathrm{Gpc}^{-3}$ 
$\mathrm{yr}^{-1}$ and $\mathcal{R}^\mathrm{hh}(z=0)=$ 0.0714 
$\mathrm{Gpc}^{-3}$ $\mathrm{yr}^{-1}$, respectively.
The delay time of `low mass + high mass' is so short that the merger 
rate density $\mathcal{R}^\mathrm{lh}(z)$ is 0 where $z\lesssim3$.

Our merger rate density of BBHs beyond the mass gap is higher than the 
upper limit obtained from GW observations.
We discuss this disagreement in the next section.

\section{Discussion}

\begin{figure}
   \begin{center}
      \includegraphics[width=\columnwidth]{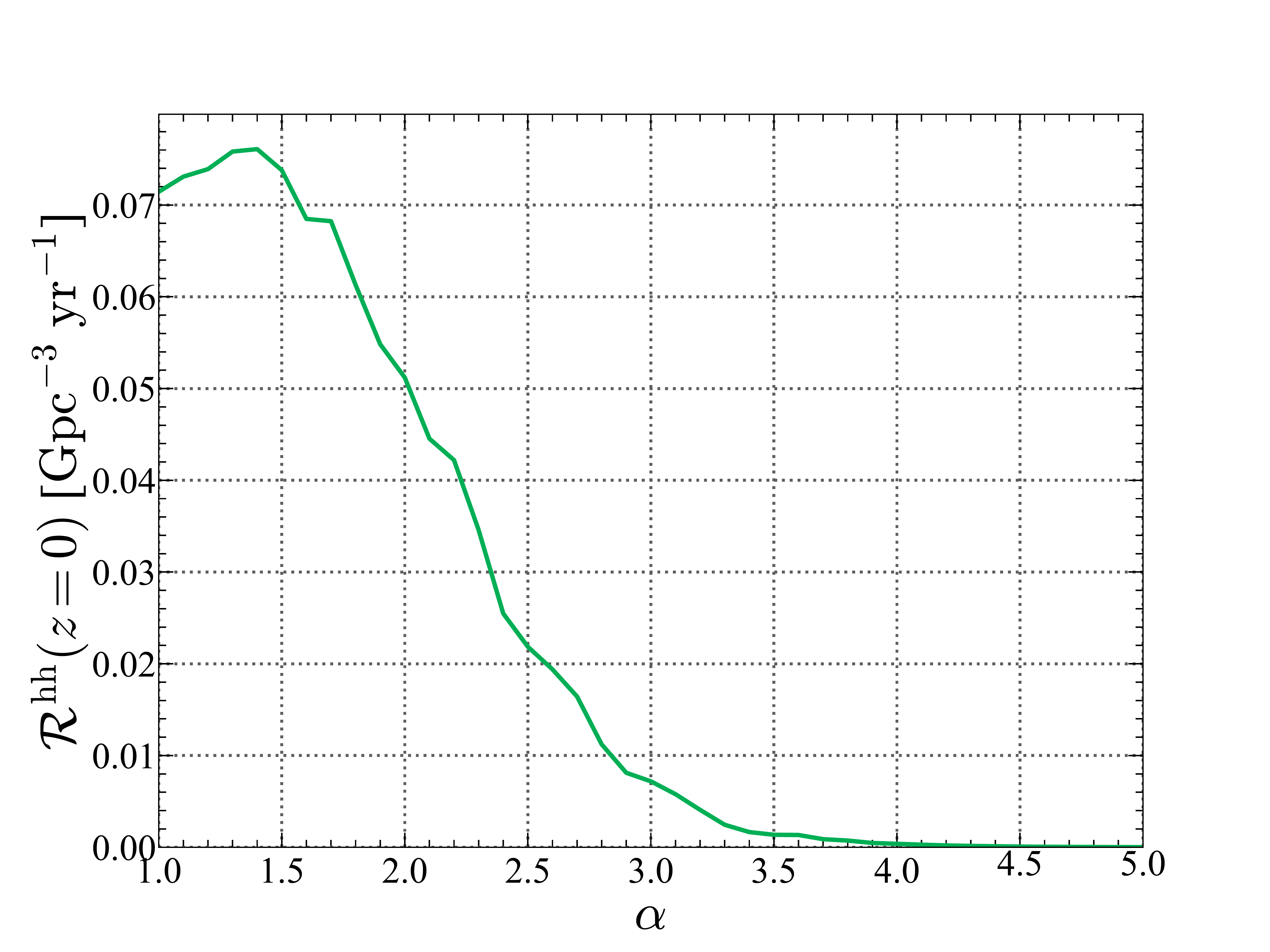}
   \end{center}
   \caption{Dependence of $\mathcal{R}^\mathrm{hh}(z=0)$ on the exponent 
   of IMF $\alpha$.
	\label{fig:a-rate_hh_z0}}
\end{figure}

So far, no BBH merger beyond the pair-instability mass gap has been
observed \citep[but see][]{fishbach20, nitz21}.
According to \citet{mariaEzquiaga20}, the current GW detectors can 
observe the BBH mergers beyond the mass gap if they exist within their 
detection horizons.
In the absence of detections, they also set the upper limit on the 
merger rate density above the mass gap at $z=0$ to be 0.01
$\mathrm{Gpc}^{-3}$ $\mathrm{yr}^{-1}$.
Our merger rate density of `high mass + high mass' 
$\mathcal{R}^\mathrm{hh}(z=0)$ is 0.0714 
$\mathrm{Gpc}^{-3}$ $\mathrm{yr}^{-1}$ and is higher than this upper 
limit.
The slope of the Pop. III IMF may be responsible for this discrepancy.
Although the Pop. III IMF is not well understood, we adopt the 
logarithmically flat IMF ($\xi_m(m_\mathrm{ZAMS,p}) \propto m_\mathrm{ZAMS,p}^{-1}$) in this study.
In order to decrease $\mathcal{R}^\mathrm{hh}(z=0)$, the Pop. III IMF 
needs to be steeper or have a cutoff.
Here, we investigate how $\mathcal{R}^\mathrm{hh}(z=0)$ would change if 
a different IMF is adopted.
We simply assume that the Pop. III IMF is expressed as a single power 
law, i.e., $\xi_m(m_\mathrm{ZAMS,p}) \propto m_\mathrm{ZAMS,p}^{-\alpha}$.
Figure \ref{fig:a-rate_hh_z0} shows the dependence of 
$\mathcal{R}^\mathrm{hh}(z=0)$ on the exponent of IMF $\alpha$.
In order for $\mathcal{R}^\mathrm{hh}(z=0)$ not to exceed the upper 
limit, 0.01 $\mathrm{Gpc}^{-3}$ $\mathrm{yr}^{-1}$, $\alpha$ must be 
greater than 2.8.
\revise{Note that, the Pop. III star formation rate and common envelope
parameters we adopt may affect our result. For example, the Pop. III star
formation rate is lower, less steeper IMFs (smaller $\alpha$) are 
permissible.}

\begin{figure}
   \begin{center}
      \includegraphics[width=\columnwidth]{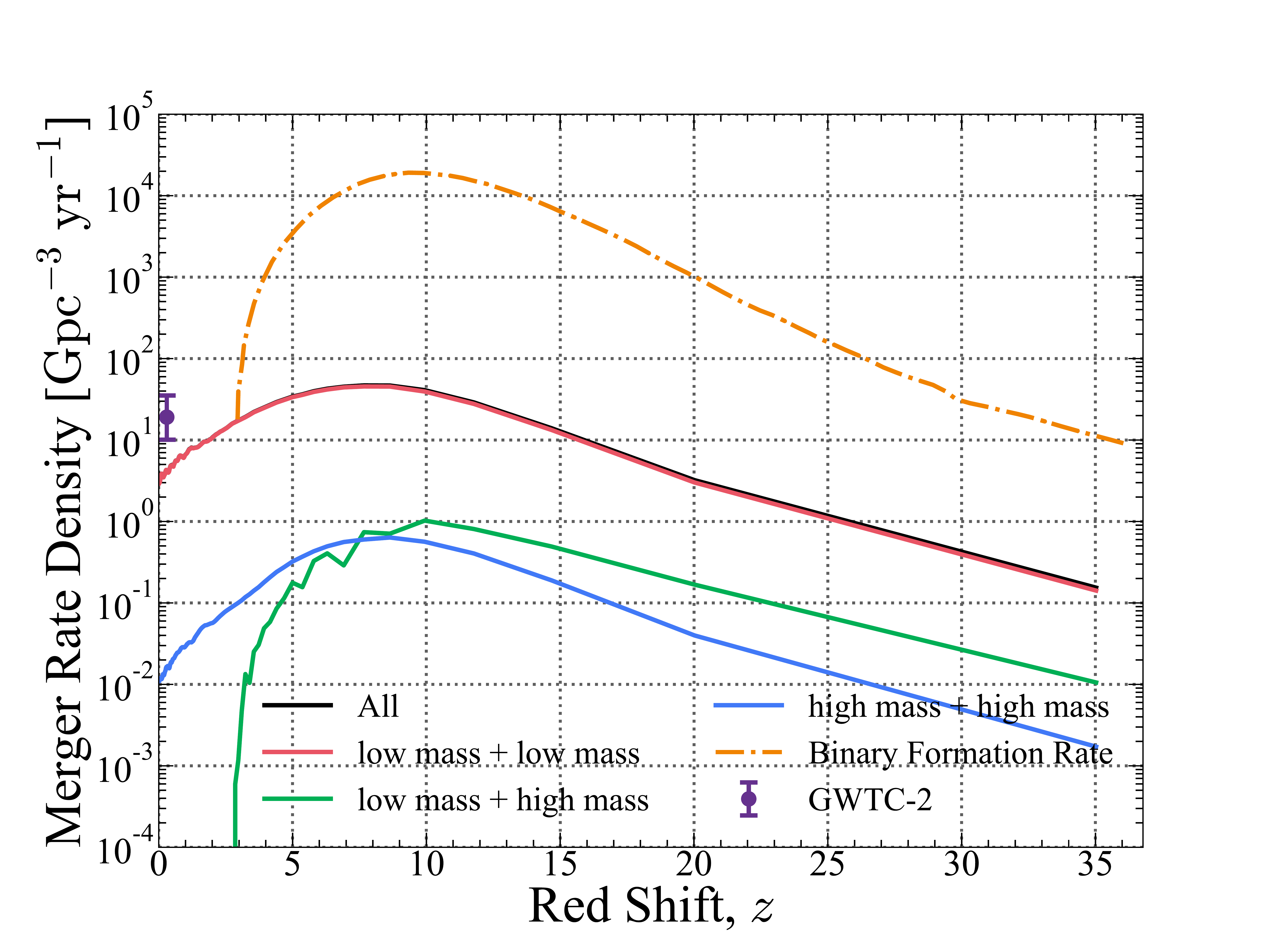}
   \end{center}
   \caption{Redshift evolution of merger rate density when $\alpha=2.8$.
   The color codes and line styles are the same as Figure
   \ref{fig:mergerrate}.
	\label{fig:mergerrate_a}}
\end{figure}

We also obtain the redshift evolution of the merger rate density when
$\alpha=2.8$ (Figure \ref{fig:mergerrate_a}).
The merger rate density of all Pop. III BBHs evolves with redshift 
according to $(1+z)^{1.21}$ within $z<2$, and the local merger rate 
density of all the BBHs $\mathcal{R}^\mathrm{all}(z=0)$ is 2.89 
$\mathrm{Gpc}^{-3}$ $\mathrm{yr}^{-1}$.
The merger rate density peaks at around $z\sim8$, and this reflects the
peak of star formation rate.
\citet{kinugawa20} also performed population synthesis calculations and
derived that the merger rate density of Pop. III BBH mergers at $z=0$ is 
3.34--21.2 $\mathrm{Gpc}^{-3}$ $\mathrm{yr}^{-1}$.
Our merger rate density is roughly consistent with theirs.
The BBH merger rate density estimated by \citet{abbott20gwtc2Property}
from GW observations is $19.1^{+16.2}_{-9.0}$ $\mathrm{Gpc}^{-3}$ 
$\mathrm{yr}^{-1}$ at $z=0$.
Thus, some portions of BBHs observed by aLIGO and Virgo may be Pop. III
origin.

\begin{figure}
   \begin{center}
      \includegraphics[width=\columnwidth]{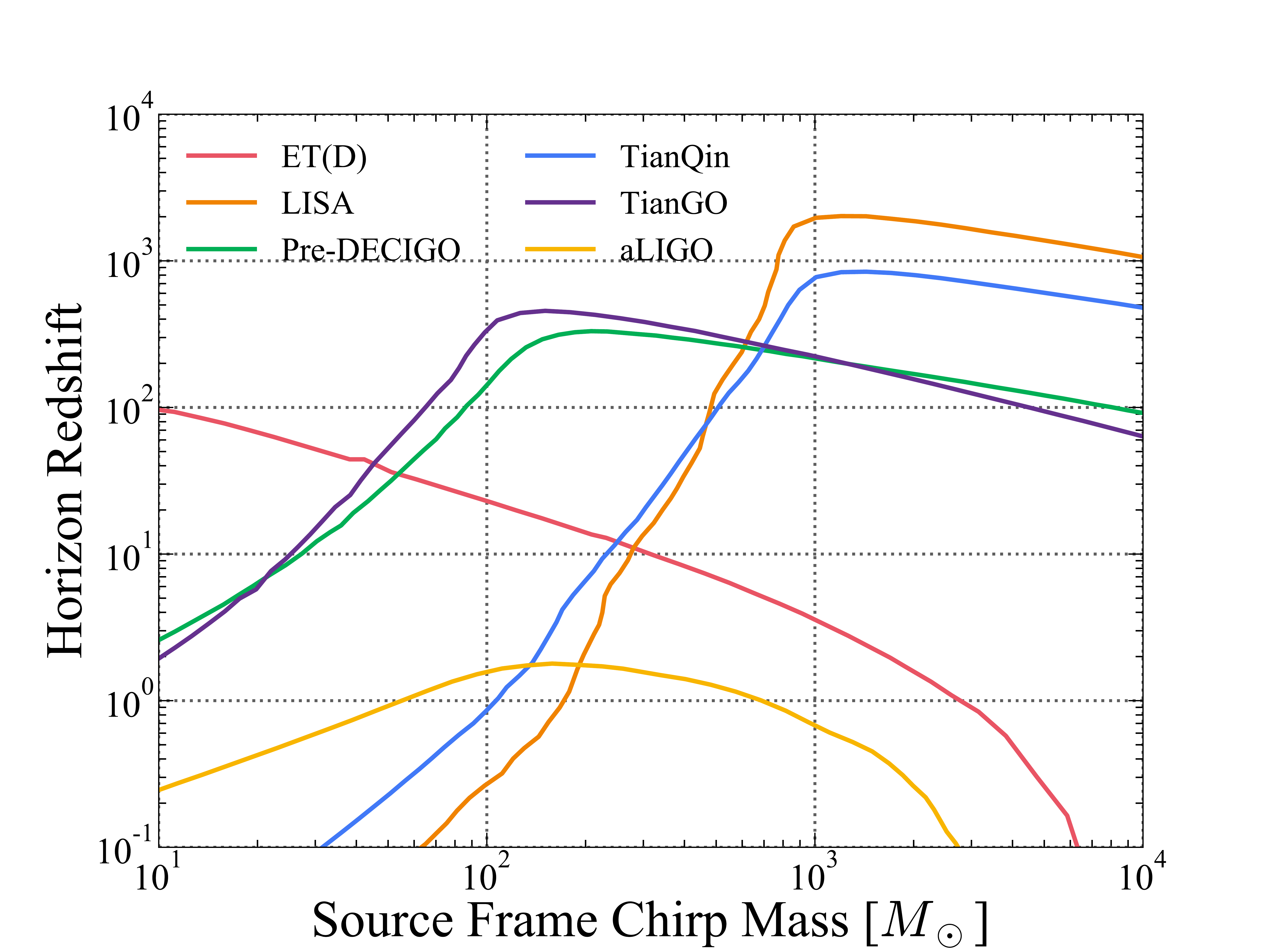}
   \end{center}
   \caption{Horizon redshift of GW detectors for %equal mass and 
   face-on compact binary mergers. The red, orange, green, blue, purple 
   and yellow solid lines indicates the horizons of Einstein telescope, 
   LISA, Pre-DECIGO, TianQin, TianGO and aLIGO.
   \label{fig:horizon_z}}
\end{figure}

\begin{table}
   \caption{Detection rate of \revise{Pop.III} BBH mergers beyond the mass gap when 
   $\alpha=2.8$.}
   \label{table:detection_rate}
   \begin{center}
     \begin{tabular}{cc}\hline
      Observatory & Detection rate [$\mathrm{yr}^{-1}$]\\
      \hline
      Einstein telescope & 126.1\\
      LISA & 1.1\\
      Pre-DECIGO & 200.9\\
      TianGO & 200.9\\
      TianQin & 7.9\\
      \hline
     \end{tabular}
  \end{center}
\end{table}
 
In the future, the third generation ground-based GW observatories,
such as Einstein telescope \citep{punturo10,sathyaprakash2012}, and
space-borne GW observatories, such as LISA \citep{amaro-seoane2017},
Pre-DECIGO \citep{kawamura2006,nakamura2016}, TianQin \citep{luo2016,
wang2019}, and TianGO \citep{kuns2020} will operate.
The detection horizons of these detectors and aLIGO are shown in Figure 
\ref{fig:horizon_z}.
Compared to aLIGO, these observatories will detect many compact binary 
mergers, and will observe BBHs beyond the mass gap up to $z\sim$10--1000.
Furthermore, by comparing the merger rate density obtained from 
observation with that derived theoretically, we may be able to impose a 
stringent limit on the Pop. III IMF.
Here, assuming $\alpha=2.8$, we estimate the detection rate of BBHs 
beyond the mass gap for some GW detectors, and the results are 
summarized in Table \ref{table:detection_rate}.
%For simplicity, we assume that all BBHs are equal mass binaries.
In our calculation, most of the BBHs beyond the mass gap have the chirp 
mass of 100--300 $\msun$.
Pre-DECIGO and TianGO can see beyond the beginnings of Pop. III star 
formation in this mass range, and the detection rate is as high as 200.9
$\mathrm{yr}^{-1}$.
%90 \% of BBHs beyond the mass gap lies within 
%$m_\mathrm{t}=$300--500 $\msun$, where $m_\mathrm{t}$ is the total 
%source frame mass.
The horizon redshift of LISA in this mass range is lower than that of
the others, and thus the detection rate is also lower.
Since $\alpha\gtrsim2.8$ is needed as discussed above, the detection 
rates in Table \ref{table:detection_rate} can also be interpreted as 
upper limits.

\begin{figure}
   \begin{center}
      \includegraphics[width=\columnwidth]{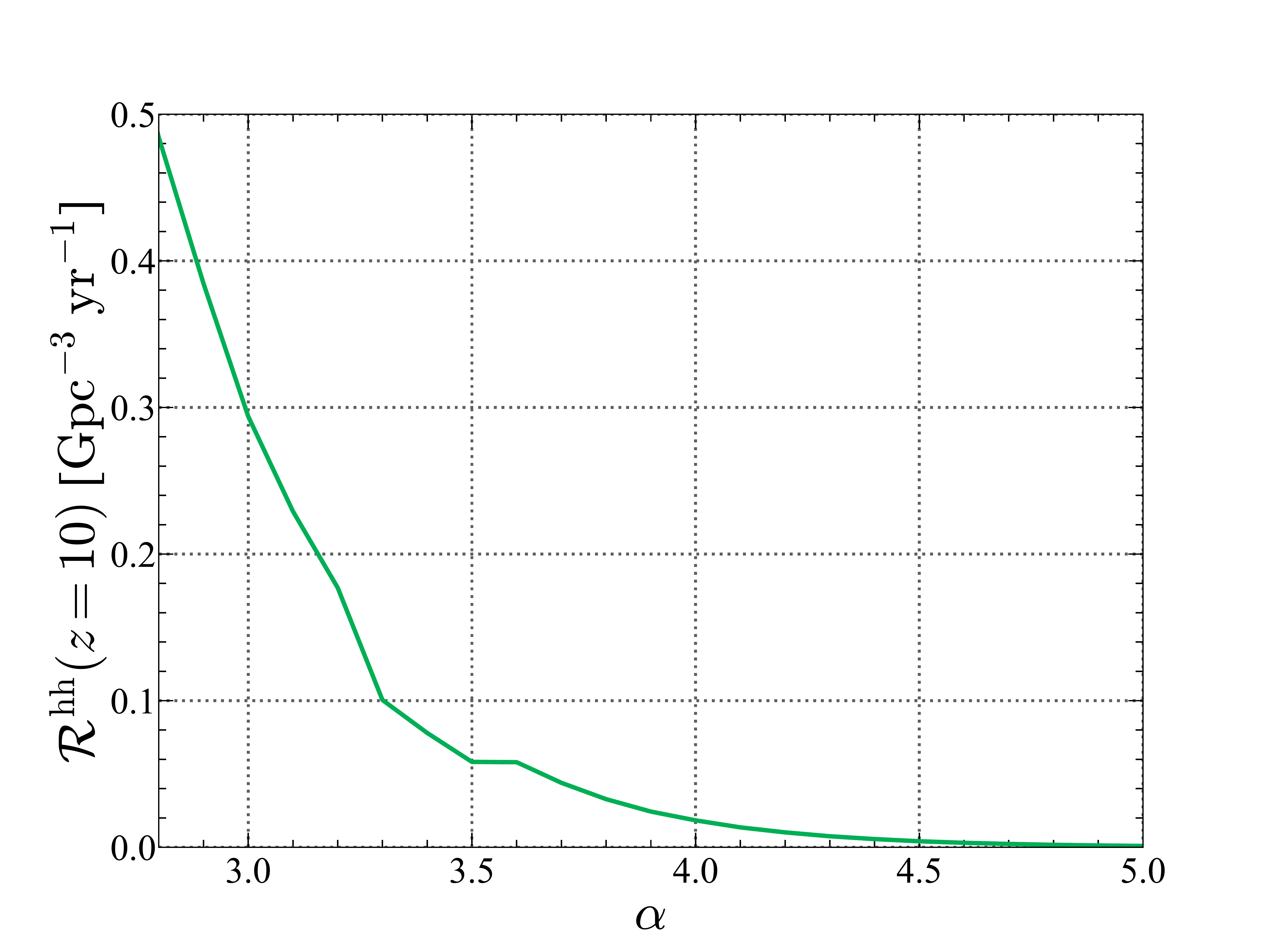}
   \end{center}
   \caption{Relation between the merger rate density of BBHs beyond the 
   mass gap at $z=10$ $\mathcal{R}^\mathrm{hh}(z=10)$ and the exponent 
   of Pop. III IMF $\alpha$.
	\label{fig:a_z10}}
\end{figure}

Let us consider that future GW detectors observe BBH
mergers, and the merger rate density at $z=10$ is obtained.
We investigate how the exponent $\alpha$ is restricted from the merger
rate density of BBHs beyond the mass gap at $z=10$,
$\mathcal{R}^\mathrm{hh}(z=10)$ (see Figure \ref{fig:a_z10}).
For example, if $\mathcal{R}^\mathrm{hh}(z=10)=$ 0.1--0.3
$\mathrm{Gpc}^{-3}$ $\mathrm{yr}^{-1}$, the exponent of IMF is
restricted to 3.0--3.3.
In this way, we may be able to impose a stringent restrictions on the 
Pop. III IMF by future GW observations.
Note that we assume that all BBH mergers beyond the mass gap at $z=10$
are originating from Pop. III stars.
In this Letter, we consider only the single power law IMF for Pop. III 
stars.
However, a double power law IMF or an IMF with a cutoff are also 
possible candidates for the Pop. III IMF.
In the future study, we will adopt such an IMF and investigate how we 
can impose restrictions on Pop. III IMF from GW observations.

\section{Summary}
In this Letter, we performed a binary population synthesis incorporating 
very massive Pop. III stars up to 1500 $\msun$ and investigated the 
sub-populations beyond the pair-instability mass gap.
Our main results are as follows.

The typical primary BH mass of `high mass + high mass' is 135--340 
$\msun$, and the maximum primary BH mass is as high as 686 $\msun$.
%The secondary BH mass of `low mass + high mass' is 45 $\msun$.
$\chi_\mathrm{eff}$ of `high mass + high mass' peaks at around 0, and
99.8 \% of those have $\chi_\mathrm{eff}\lesssim0.1$.
In order for the merger rate density of BBHs beyond the mass gap not to
exceed the upper limit \citep{mariaEzquiaga20}, the exponent $\alpha$ of
single power law IMF needs to be greater than 2.8.
When $\alpha=2.8$, space-borne GW observatories, Pre-DECIGO and TianGO, 
will be able to detect as much as $\sim200$ BBHs beyond the mass gap per
year.
We suggest that we may be able to impose a stringent limit on the Pop. 
III IMF by comparing the merger rate density obtained from future 
observations with that derived theoretically.

\section*{Acknowledgements}
% Entry for the table of contents, for this guide only
%\addcontentsline{toc}{section}{Acknowledgements}
This work was supported by JSPS KAKENHI grant No. 17H06360 and 19K03907
(A. T.), 21K13915 (T. K), 20H05249 (T. Y.), and 17H01130 and 17K05380 
(H. U.), and by the University of Tokyo Young Excellent Researcher 
program (T. K.).
We use the Python packages, {\tt numpy} \citep{vanderwalt11} and 
{\tt matplotlib} \citep{hunter07} for our analysis.

%%%%%%%%%%%%%%%%%%%%%%%%%%%%%%%%%%%%%%%%%%%%%%%%%%

\section*{Data Availability}
Results will be shared on reasonable request to authors.



\bibliographystyle{mnras}
\bibliography{natbib.bib} % if your bibtex file is called example.bib

\begin{thebibliography}{}
\makeatletter
\relax
\def\mn@urlcharsother{\let\do\@makeother \do\$\do\&\do\#\do\^\do\_\do\%\do\~}
\def\mn@doi{\begingroup\mn@urlcharsother \@ifnextchar [ {\mn@doi@}
  {\mn@doi@[]}}
\def\mn@doi@[#1]#2{\def\@tempa{#1}\ifx\@tempa\@empty \href
  {http://dx.doi.org/#2} {doi:#2}\else \href {http://dx.doi.org/#2} {#1}\fi
  \endgroup}
\def\mn@eprint#1#2{\mn@eprint@#1:#2::\@nil}
\def\mn@eprint@arXiv#1{\href {http://arxiv.org/abs/#1} {{\tt arXiv:#1}}}
\def\mn@eprint@dblp#1{\href {http://dblp.uni-trier.de/rec/bibtex/#1.xml}
  {dblp:#1}}
\def\mn@eprint@#1:#2:#3:#4\@nil{\def\@tempa {#1}\def\@tempb {#2}\def\@tempc
  {#3}\ifx \@tempc \@empty \let \@tempc \@tempb \let \@tempb \@tempa \fi \ifx
  \@tempb \@empty \def\@tempb {arXiv}\fi \@ifundefined
  {mn@eprint@\@tempb}{\@tempb:\@tempc}{\expandafter \expandafter \csname
  mn@eprint@\@tempb\endcsname \expandafter{\@tempc}}}

\bibitem[\protect\citeauthoryear{{Abbott} et~al.,}{{Abbott}
  et~al.}{2020}]{abbott20gwtc2}
{Abbott} R.,  et~al., 2020, arXiv e-prints, \href
  {https://ui.adsabs.harvard.edu/abs/2020arXiv201014527A} {p. arXiv:2010.14527}

\bibitem[\protect\citeauthoryear{{Amaro-Seoane} et~al.,}{{Amaro-Seoane}
  et~al.}{2017}]{amaro-seoane2017}
{Amaro-Seoane} P.,  et~al., 2017, arXiv e-prints, \href
  {https://ui.adsabs.harvard.edu/abs/2017arXiv170200786A} {p. arXiv:1702.00786}

\bibitem[\protect\citeauthoryear{{Antonini}, {Toonen}  \& {Hamers}}{{Antonini}
  et~al.}{2017}]{antonini17}
{Antonini} F.,  {Toonen} S.,   {Hamers} A.~S.,  2017, \mn@doi [\apj]
  {10.3847/1538-4357/aa6f5e}, \href
  {https://ui.adsabs.harvard.edu/abs/2017ApJ...841...77A} {841, 77}

\bibitem[\protect\citeauthoryear{{Belczynski}, {Bulik}  \&
  {Rudak}}{{Belczynski} et~al.}{2004}]{belczynski04}
{Belczynski} K.,  {Bulik} T.,   {Rudak} B.,  2004, \mn@doi [\apjl]
  {10.1086/422172}, \href
  {https://ui.adsabs.harvard.edu/abs/2004ApJ...608L..45B} {608, L45}

\bibitem[\protect\citeauthoryear{{Belczynski} et~al.,}{{Belczynski}
  et~al.}{2016}]{belczynski16b}
{Belczynski} K.,  et~al., 2016, \mn@doi [\aap] {10.1051/0004-6361/201628980},
  \href {https://ui.adsabs.harvard.edu/abs/2016A&A...594A..97B} {594, A97}

\bibitem[\protect\citeauthoryear{{Belczynski} et~al.,}{{Belczynski}
  et~al.}{2020}]{belczynski20}
{Belczynski} K.,  et~al., 2020, \mn@doi [\aap] {10.1051/0004-6361/201936528},
  \href {https://ui.adsabs.harvard.edu/abs/2020A&A...636A.104B} {636, A104}

\bibitem[\protect\citeauthoryear{{Di Carlo} et~al.,}{{Di Carlo}
  et~al.}{2020}]{dicarlo20}
{Di Carlo} U.~N.,  et~al., 2020, \mn@doi [\mnras] {10.1093/mnras/staa2286},
  \href {https://ui.adsabs.harvard.edu/abs/2020MNRAS.498..495D} {498, 495}

\bibitem[\protect\citeauthoryear{{Fishbach} \& {Holz}}{{Fishbach} \&
  {Holz}}{2020}]{fishbach20}
{Fishbach} M.,  {Holz} D.~E.,  2020, \mn@doi [\apjl]
  {10.3847/2041-8213/abc827}, \href
  {https://ui.adsabs.harvard.edu/abs/2020ApJ...904L..26F} {904, L26}

\bibitem[\protect\citeauthoryear{{Fryer}, {Belczynski}, {Wiktorowicz},
  {Dominik}, {Kalogera}  \& {Holz}}{{Fryer} et~al.}{2012}]{fryer12}
{Fryer} C.~L.,  {Belczynski} K.,  {Wiktorowicz} G.,  {Dominik} M.,  {Kalogera}
  V.,   {Holz} D.~E.,  2012, \mn@doi [\apj] {10.1088/0004-637X/749/1/91}, \href
  {http://adsabs.harvard.edu/abs/2012ApJ...749...91F} {749, 91}

\bibitem[\protect\citeauthoryear{{Hirano}, {Hosokawa}, {Yoshida}, {Omukai}  \&
  {Yorke}}{{Hirano} et~al.}{2015}]{hirano15}
{Hirano} S.,  {Hosokawa} T.,  {Yoshida} N.,  {Omukai} K.,   {Yorke} H.~W.,
  2015, \mn@doi [\mnras] {10.1093/mnras/stv044}, \href
  {https://ui.adsabs.harvard.edu/abs/2015MNRAS.448..568H} {448, 568}

\bibitem[\protect\citeauthoryear{{Hunter}}{{Hunter}}{2007}]{hunter07}
{Hunter} J.~D.,  2007, \mn@doi [Computing in Science and Engineering]
  {10.1109/MCSE.2007.55}, \href
  {https://ui.adsabs.harvard.edu/abs/2007CSE.....9...90H} {9, 90}

\bibitem[\protect\citeauthoryear{{Hurley}, {Pols}  \& {Tout}}{{Hurley}
  et~al.}{2000}]{hurley00}
{Hurley} J.~R.,  {Pols} O.~R.,   {Tout} C.~A.,  2000, \mn@doi [\mnras]
  {10.1046/j.1365-8711.2000.03426.x}, \href
  {https://ui.adsabs.harvard.edu/abs/2000MNRAS.315..543H} {315, 543}

\bibitem[\protect\citeauthoryear{{Hurley}, {Tout}  \& {Pols}}{{Hurley}
  et~al.}{2002}]{hurley02}
{Hurley} J.~R.,  {Tout} C.~A.,   {Pols} O.~R.,  2002, \mn@doi [\mnras]
  {10.1046/j.1365-8711.2002.05038.x}, \href
  {https://ui.adsabs.harvard.edu/abs/2002MNRAS.329..897H} {329, 897}

\bibitem[\protect\citeauthoryear{{Inayoshi}, {Kashiyama}, {Visbal}  \&
  {Haiman}}{{Inayoshi} et~al.}{2016}]{inayoshi16}
{Inayoshi} K.,  {Kashiyama} K.,  {Visbal} E.,   {Haiman} Z.,  2016, \mn@doi
  [\mnras] {10.1093/mnras/stw1431}, \href
  {https://ui.adsabs.harvard.edu/abs/2016MNRAS.461.2722I} {461, 2722}

\bibitem[\protect\citeauthoryear{{Kawamura} et~al.,}{{Kawamura}
  et~al.}{2006}]{kawamura2006}
{Kawamura} S.,  et~al., 2006, \mn@doi [Classical and Quantum Gravity]
  {10.1088/0264-9381/23/8/S17}, \href
  {https://ui.adsabs.harvard.edu/abs/2006CQGra..23S.125K} {23, S125}

\bibitem[\protect\citeauthoryear{{Kinugawa}, {Inayoshi}, {Hotokezaka},
  {Nakauchi}  \& {Nakamura}}{{Kinugawa} et~al.}{2014}]{kinugawa14}
{Kinugawa} T.,  {Inayoshi} K.,  {Hotokezaka} K.,  {Nakauchi} D.,   {Nakamura}
  T.,  2014, \mn@doi [\mnras] {10.1093/mnras/stu1022}, \href
  {https://ui.adsabs.harvard.edu/abs/2014MNRAS.442.2963K} {442, 2963}

\bibitem[\protect\citeauthoryear{{Kinugawa}, {Nakamura}  \&
  {Nakano}}{{Kinugawa} et~al.}{2020}]{kinugawa20}
{Kinugawa} T.,  {Nakamura} T.,   {Nakano} H.,  2020, \mn@doi [\mnras]
  {10.1093/mnras/staa2511}, \href
  {https://ui.adsabs.harvard.edu/abs/2020MNRAS.498.3946K} {498, 3946}

\bibitem[\protect\citeauthoryear{{Kinugawa}, {Nakamura}  \&
  {Nakano}}{{Kinugawa} et~al.}{2021}]{kinugawa21}
{Kinugawa} T.,  {Nakamura} T.,   {Nakano} H.,  2021, \mn@doi [\mnras]
  {10.1093/mnrasl/slab032}, \href
  {https://ui.adsabs.harvard.edu/abs/2021MNRAS.504L..28K} {504, L28}

\bibitem[\protect\citeauthoryear{{Kruckow}, {Tauris}, {Langer}, {Kramer}  \&
  {Izzard}}{{Kruckow} et~al.}{2018}]{kruckow18}
{Kruckow} M.~U.,  {Tauris} T.~M.,  {Langer} N.,  {Kramer} M.,   {Izzard} R.~G.,
   2018, \mn@doi [\mnras] {10.1093/mnras/sty2190}, \href
  {http://ads.nao.ac.jp/abs/2018MNRAS.481.1908K} {481, 1908}

\bibitem[\protect\citeauthoryear{{Kumamoto}, {Fujii}  \& {Tanikawa}}{{Kumamoto}
  et~al.}{2020}]{kumamoto20}
{Kumamoto} J.,  {Fujii} M.~S.,   {Tanikawa} A.,  2020, \mn@doi [\mnras]
  {10.1093/mnras/staa1440}, \href
  {https://ui.adsabs.harvard.edu/abs/2020MNRAS.495.4268K} {495, 4268}

\bibitem[\protect\citeauthoryear{{Kuns}, {Yu}, {Chen}  \& {Adhikari}}{{Kuns}
  et~al.}{2020}]{kuns2020}
{Kuns} K.~A.,  {Yu} H.,  {Chen} Y.,   {Adhikari} R.~X.,  2020, \mn@doi [\prd]
  {10.1103/PhysRevD.102.043001}, \href
  {https://ui.adsabs.harvard.edu/abs/2020PhRvD.102d3001K} {102, 043001}

\bibitem[\protect\citeauthoryear{{Luo} et~al.,}{{Luo} et~al.}{2016}]{luo2016}
{Luo} J.,  et~al., 2016, \mn@doi [Classical and Quantum Gravity]
  {10.1088/0264-9381/33/3/035010}, \href
  {https://ui.adsabs.harvard.edu/abs/2016CQGra..33c5010L} {33, 035010}

\bibitem[\protect\citeauthoryear{{Mar{\'\i}a Ezquiaga} \& {Holz}}{{Mar{\'\i}a
  Ezquiaga} \& {Holz}}{2020}]{mariaEzquiaga20}
{Mar{\'\i}a Ezquiaga} J.,  {Holz} D.~E.,  2020, arXiv e-prints, \href
  {https://ui.adsabs.harvard.edu/abs/2020arXiv200602211M} {p. arXiv:2006.02211}

\bibitem[\protect\citeauthoryear{{Nakamura} et~al.,}{{Nakamura}
  et~al.}{2016}]{nakamura2016}
{Nakamura} T.,  et~al., 2016, \mn@doi [Progress of Theoretical and Experimental
  Physics] {10.1093/ptep/ptw170}, \href
  {https://ui.adsabs.harvard.edu/abs/2016PTEP.2016l9301N} {2016, 129301}

\bibitem[\protect\citeauthoryear{{Nitz} \& {Capano}}{{Nitz} \&
  {Capano}}{2021}]{nitz21}
{Nitz} A.~H.,  {Capano} C.~D.,  2021, \mn@doi [\apjl]
  {10.3847/2041-8213/abccc5}, \href
  {https://ui.adsabs.harvard.edu/abs/2021ApJ...907L...9N} {907, L9}

\bibitem[\protect\citeauthoryear{{Punturo} et~al.,}{{Punturo}
  et~al.}{2010}]{punturo10}
{Punturo} M.,  et~al., 2010, \mn@doi [Classical and Quantum Gravity]
  {10.1088/0264-9381/27/19/194002}, \href
  {https://ui.adsabs.harvard.edu/abs/2010CQGra..27s4002P} {27, 194002}

\bibitem[\protect\citeauthoryear{{Reitze} et~al.,}{{Reitze}
  et~al.}{2019}]{reitze19}
{Reitze} D.,  et~al., 2019, in Bulletin of the American Astronomical Society.
  p.~35 (\mn@eprint {arXiv} {1907.04833})

\bibitem[\protect\citeauthoryear{{Rodriguez}, {Chatterjee}  \&
  {Rasio}}{{Rodriguez} et~al.}{2016}]{rodriguez16}
{Rodriguez} C.~L.,  {Chatterjee} S.,   {Rasio} F.~A.,  2016, \mn@doi [\prd]
  {10.1103/PhysRevD.93.084029}, \href
  {https://ui.adsabs.harvard.edu/abs/2016PhRvD..93h4029R} {93, 084029}

\bibitem[\protect\citeauthoryear{{Sathyaprakash} et~al.,}{{Sathyaprakash}
  et~al.}{2012}]{sathyaprakash2012}
{Sathyaprakash} B.,  et~al., 2012, \mn@doi [Classical and Quantum Gravity]
  {10.1088/0264-9381/29/12/124013}, \href
  {https://ui.adsabs.harvard.edu/abs/2012CQGra..29l4013S} {29, 124013}

\bibitem[\protect\citeauthoryear{{Susa}, {Hasegawa}  \& {Tominaga}}{{Susa}
  et~al.}{2014}]{susa14}
{Susa} H.,  {Hasegawa} K.,   {Tominaga} N.,  2014, \mn@doi [\apj]
  {10.1088/0004-637X/792/1/32}, \href
  {https://ui.adsabs.harvard.edu/abs/2014ApJ...792...32S} {792, 32}

\bibitem[\protect\citeauthoryear{{Tagawa}, {Haiman}  \& {Kocsis}}{{Tagawa}
  et~al.}{2020}]{tagawa20}
{Tagawa} H.,  {Haiman} Z.,   {Kocsis} B.,  2020, \mn@doi [\apj]
  {10.3847/1538-4357/ab9b8c}, \href
  {https://ui.adsabs.harvard.edu/abs/2020ApJ...898...25T} {898, 25}

\bibitem[\protect\citeauthoryear{{Tanikawa}, {Yoshida}, {Kinugawa}, {Takahashi}
   \& {Umeda}}{{Tanikawa} et~al.}{2020}]{tanikawa20a}
{Tanikawa} A.,  {Yoshida} T.,  {Kinugawa} T.,  {Takahashi} K.,   {Umeda} H.,
  2020, \mn@doi [\mnras] {10.1093/mnras/staa1417}, \href
  {https://ui.adsabs.harvard.edu/abs/2020MNRAS.495.4170T} {495, 4170}

\bibitem[\protect\citeauthoryear{{Tanikawa}, {Susa}, {Yoshida}, {Trani}  \&
  {Kinugawa}}{{Tanikawa} et~al.}{2021}]{tanikawa21}
{Tanikawa} A.,  {Susa} H.,  {Yoshida} T.,  {Trani} A.~A.,   {Kinugawa} T.,
  2021, \mn@doi [\apj] {10.3847/1538-4357/abe40d}, \href
  {https://ui.adsabs.harvard.edu/abs/2021ApJ...910...30T} {910, 30}

\bibitem[\protect\citeauthoryear{{The LIGO Scientific Collaboration}
  et~al.,}{{The LIGO Scientific Collaboration}
  et~al.}{2020}]{abbott20gwtc2Property}
{The LIGO Scientific Collaboration} et~al., 2020, arXiv e-prints, \href
  {https://ui.adsabs.harvard.edu/abs/2020arXiv201014533T} {p. arXiv:2010.14533}

\bibitem[\protect\citeauthoryear{{Wang} et~al.,}{{Wang}
  et~al.}{2019}]{wang2019}
{Wang} H.-T.,  et~al., 2019, \mn@doi [\prd] {10.1103/PhysRevD.100.043003},
  \href {https://ui.adsabs.harvard.edu/abs/2019PhRvD.100d3003W} {100, 043003}

\bibitem[\protect\citeauthoryear{{Webbink}}{{Webbink}}{1984}]{webbink84}
{Webbink} R.~F.,  1984, \mn@doi [\apj] {10.1086/161701}, \href
  {http://adsabs.harvard.edu/abs/1984ApJ...277..355W} {277, 355}

\bibitem[\protect\citeauthoryear{{de Souza}, {Yoshida}  \& {Ioka}}{{de Souza}
  et~al.}{2011}]{desouza11}
{de Souza} R.~S.,  {Yoshida} N.,   {Ioka} K.,  2011, \mn@doi [\aap]
  {10.1051/0004-6361/201117242}, \href
  {https://ui.adsabs.harvard.edu/abs/2011A&A...533A..32D} {533, A32}

\bibitem[\protect\citeauthoryear{{van der Walt}, {Colbert}  \&
  {Varoquaux}}{{van der Walt} et~al.}{2011}]{vanderwalt11}
{van der Walt} S.,  {Colbert} S.~C.,   {Varoquaux} G.,  2011, \mn@doi
  [Computing in Science and Engineering] {10.1109/MCSE.2011.37}, \href
  {https://ui.adsabs.harvard.edu/abs/2011CSE....13b..22V} {13, 22}

\makeatother
\end{thebibliography}


%%%%%%%%%%%%%%%%%%%%%%%%%%%%%%%%%%%%%%%%%%%%%%%%%%

%%%%%%%%%%%%%%%%% APPENDICES %%%%%%%%%%%%%%%%%%%%%

%\appendix
%\section{Journal abbreviations}

%%%%%%%%%%%%%%%%%%%%%%%%%%%%%%%%%%%%%%%%%%%%%%%%%%

% Don't change these lines
\bsp	% typesetting comment
\label{lastpage}
\end{document}